\documentclass[letterpaper,prb,preprint]{revtex4}
\usepackage{natbib}
\usepackage{epsfig}
\usepackage{subfigure}
\usepackage{mhchem}
\usepackage{gensymb}
\usepackage{cleveref}
\usepackage{graphicx}
\usepackage{amsmath}
\usepackage{chemmacros}
\usepackage{rotating}
\usepackage{bm}
\begin{document}
\title{First Principles Investigation of Transition Metal Doped WSe$_{2}$ Monolayer for Photocatalytic Water Splitting}
\author{Celine Wu}
\author{Xuan Luo}
\affiliation{ National Graphene Research and Development Center, Springfield, Virginia 22151, USA}
\date{\today}

\newpage
\maketitle
\section{Introduction}
Fossil fuels have been the primary source of energy for the past two centuries\cite{EIA2011}. However, the growing concern about the use of fossil fuels has drastically increased as its consumption has exceedingly outpaced the amount of resources. In 2018, the United States' energy consumption hit a record high of 101.3 quadrillion British thermal units(Btu)\cite{EIA2019}, among which, 80\% were fossil fuels\cite{DesSilver2020, EIA2020}. It is predicted that oil will be exhausted in 30 years, gas in 40 years, and coal in 70 years\cite{Kuo2019}. Consequently, there is a pressing need for renewable energy sources. Recently, photocatalysis, in particular, photocatalytic water splitting, has drawn great attention as an environmentally friendly solution to this crisis by converting solar energy into hydrogen fuel\cite{GaAs2020, ecofriendly2019}.

Photocatalysis results in the rate change of chemical transformation due to the presence of a photocatalyst, a material which absorbs light and acts as a reagent, but is not consumed in the process\cite{Suib2013, Photocatalysis2018}. The key to photocatalytic water splitting is the material for the photocatalyst because it absorbs phonon energy, which excites electrons from the occupied valence band to the unoccupied conduction band\cite{Process2017}. This process generates electron–hole pairs which spread to the surface and take part in the two half-redox reactions. The reactions split the hydrogen-oxygen bond \ce[{H$_{2}$O $\rightarrow$ H$_{2}$ + O}]\cite{Process2017, Splitting2018, Reactions2017, Takanabe2017}. The hydrogen undergoes a hydrogen evolution reaction(HER), as shown in equation 1, and generates gaseous hydrogen\cite{HER2019, HERWebsite}. The oxygen undergoes oxygen evolution reaction(OER)(equation 2), which is needed for energy conversion and energy storage\cite{OER2019}. Therefore, searching for an ideal material for the photocatalyst is vital for effective photocatalytic water splitting as an alternative energy source.
\begin{equation}
\ch{2 H^{+} + 2 e^{-} ->[0eV] H_{2}}
\end{equation}
\begin{equation}
\ch{2 H_{2}O + 4 h^{+} ->[1.23eV] 4 H^{+} + O_{2}}
\end{equation}
In 1972, Fujishima and Honda used TiO$_{2}$ as a photocatalyst in semiconductor electrochemical photolysis for the first time\cite{FujishimaHonda1972}, which paved the way for future research on semiconductor photocatalysis. Thirty years later, researchers started investigating the use of cocatalysts and sacrificial reagents to improve photocatalytic abilities\cite{cocatalyst2005, sacrificialReagents2005}. Now, a variety of different materials and methods have been explored to find an ideal band gap ranging from heterostructures\cite{Heterostructure2018}, to metal and non-metal doping\cite{Doping2018}, then to nanostructures\cite{Nanostructures2020}. 

Recently, 2D materials have gained attention as their photocatalytic abilities have shown promising results. In particular, several 2D materials have been investigated for photocatalytic water splitting such as SbP$_{3}$\cite{SbP32019}, SIn$_{2}$Te\cite{SIn2Te2019}, GaS\cite{GaS2015}, and MgPSe$_{3}$\cite{GaS2015}. Monolayers show superior photocatalysis compared to bulk materials due to their maximized surface area, low electron-hole recombination rate, and broad adsorption range\cite{2D2018, ShortenDistance2019, SnN32019}. 2D materials have a large surface area available for photocatalytic reactions because they optimize the amount of surface area for a material. They also reduce the probability of electron-hole recombination because the distance that the photogenerated electrons and holes have to travel in order to reach the solid/water interface is reduced\cite{ ShortenDistance2019, SnN32019}. Several recent studies proved the advantages of 2D materials. For example, Zhang et al. recently studied the photocatalytic properties of bulk Cu$_{2}$ZnSnS$_{4}$ and monolayer Cu$_{2}$ZnSnS$_{4}$. They reported that the monolayer's band edge positions were narrower to the ideal placement and its overpotentials were lower for the two reactions than the bulk\cite{Cu2ZnSnS42020}. In addition, Garg et al. compared the photocatalytic properties of monolayer and bulk CuCl. They demonstrated that the stability of the 2D material was superior to that of the 3D material due to its favorable band edge alignments and its driving force, which compensated the reaction overpotential for OER\cite{CuCl2019}. Due to the desirable properties of 2D materials, researchers have investigated a variety of different 2D materials for photocatalysis\cite{PdSeO32018, Graphdiyne2020, C2N2017, C3N42018, Cu2ZnSnS42020, WS2}. 
 
In particular, one group of 2D materials, Transition Metal Dichalcogenides(TMDs), have received special attention for their potential as catalysts. TMDs are semiconductors in the form of MX$_{2}$, where M is a transition metal element from group IV(Ti, Zr, or Hf), group V(V, Nb, Ta) or group VI(Mo or W); and X is a chalcogen(S, Se or Te)\cite{Pan2014}. TMDs have high reactivity, reasonable stability, and a suitable band gap\cite{Lolla2020, JanusTMD2020}. The wettability of TMDs has also been reported in literature. For example, Chhowalla et al. and others have shown 2D TMDs to be chemically stable in aqueous solutions, accessible in experiments, nontoxic, and inexpensive\cite{ChhowallaStability,LiStability, WangStability,ZhangStability,ZhouStability,WuStability,FanabStability}.

Among the numerous TMDs that have been studied, MoS$_{2}$ has gained the most popularity due to its adaptable characteristics. It has been doped with many different materials\cite{Pan2014, Lolla2020, AnionMoS22017, NDoping2019}, combined into heterostructures\cite{MoS2/BiOX2019, M2CO2/MoS22020}, and used as a cocatalyst\cite{Cocatalyst2019}. However, regarding the most important property of a photocatalyst \textemdash band gap\textemdash \space Mo$_{2}$ is still not ideal. The band gap of Mo$_{2}$ is 1.8eV\cite{MoS2bandgap}, which is still considerably larger than the ideal 1.23eV band gap. 

In order to further narrow the band gap, in this paper, we studied a new TMD material, WSe$_{2}$, as a potential photocatalyst. WSe$_{2}$ has a similar structure and components to MoS$_{2}$, and is currently used in heterostructures. To our knowledge, this is the first time WSe$_{2}$ is studied as a potential photocatalyst.  We aim to determine the photocatalytic abilities of WSe$_{2}$, and whether it can be improved with Cr, Mo, Ta, and Re substitutional doping using first principles calculations.

\section{Method}

The ABINIT simulation package was used to calculate the band structure, Projected Density of States(PDOS), formation energy, and adsorption energy of pristine WSe$_{2}$ and Cr, Mo, Ta, and Re doped WSe$_{2}$.

\subsection{Computational Methods}

Our calculations were based on Density Functional Theory(DFT) in conjunction with the generalized gradient approximations(GGA) for an exchange-correlation functional, as described by Perdew-Burke-Ernzerhof(PBE)\cite{GGA-PBE2016}. Projector augmented wave(PAW) pseudopotentials were generated with the AtomPAW code\cite{AtomPAW2018}. The electronic configurations and radius cut offs for generating the PAW pseudopotentials are shown in \Cref{tab:Table1}. 

\begin{table}[h]
\caption{The electronic configurations and radius cut offs used to generate the PAW pseudopotentials}
\begin{tabular}{c c c}
Element&Electronic Configuration&Radius cut off(a.u.)\\

\hline
\hline
Chromium&[Ne] 3s$^{2}$ 3p$^{6}$ 4s$^{1}$ 3d$^{5}$&2.1\\
Selenium&[Ar 3d$^{10}$] 4s$^{2}$ 4p$^{4}$&2.2\\
Molybdenum&[Ar 3d$^{10}$] 4s$^{2}$ 4p$^{6}$ 5s$^{1}$ 4d$^{5}$&2.2\\
Tantalum&[Kr 4d$^{10}$ 4f$^{14}$] 5s$^{2}$ 5p$^{6}$ 6s$^{2}$ 5d$^{3}$&2.41\\
Tungsten&[Kr 4d$^{10}$ 4f$^{14}$] 5s$^{2}$ 5p$^{6}$ 6s$^{2}$ 5d$^{4}$&2.41\\
Rhenium&[Kr 4d$^{10}$ 4f$^{14}$] 5s$^{2}$ 5p$^{6}$ 6s$^{2}$ 5d$^{5}$&2.4
\end{tabular}
\label{tab:Table1}
\end{table}

For all materials, the kinetic energy cut off and Monkhorst-Pack k-point grids converged. The self-consistent field(SCF) iterations terminated when the difference between the current and the previous total energy was smaller than 1.0$\times$10$^{-10}$ Hartree twice in a row. The kinetic energy cut off and Monkhorst-Pack kpoint grid values were considered converged when the difference between the current and previous total energy values for all the different data sets were less than 0.0001 Hartree(about 3meV) twice in succession. All of the atomic positions were relaxed before the convergence, and the lattice parameters were relaxed before and after convergence.  We applied the Broyden-Fletcher-Goldfarb-Shanno(BFGS) minimization with a maximal force tolerance of 5.0$\times$10$^{-5}$ Hartree/Bohr. Once the maximal force tolerance was reached, the structural relaxation iterations terminated. When the difference between the current and previous total force for all atoms reached 5.0$\times$10$^{-6}$ Hartree/Bohr twice consecutively, the SCF cycle was completed. These relaxed and converged values were then used for total energy calculations.

\subsection{Electronic Structure}

\begin{figure}[h]
(a)\includegraphics[width=0.5\linewidth]{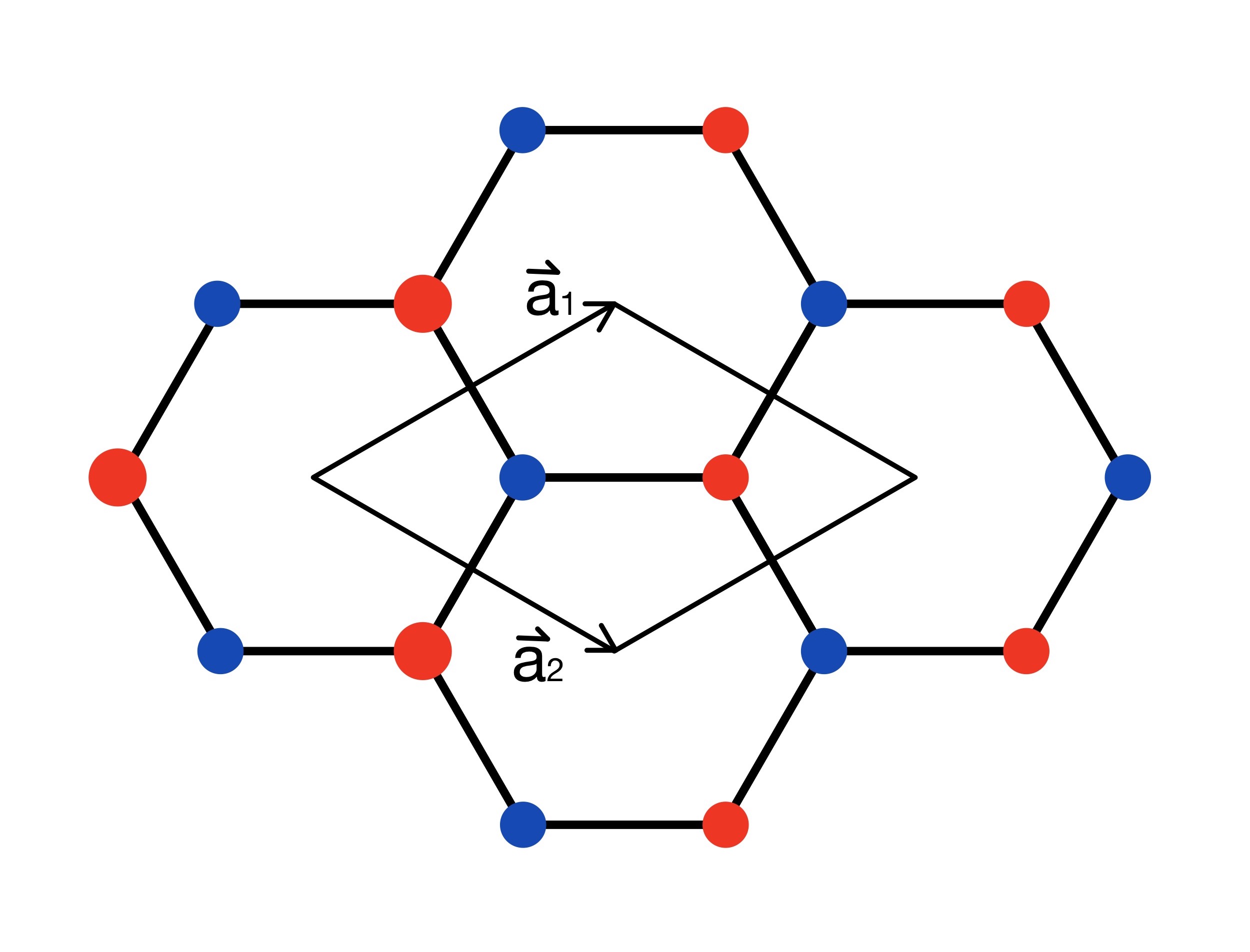}
(b)\includegraphics[width=0.35\linewidth]{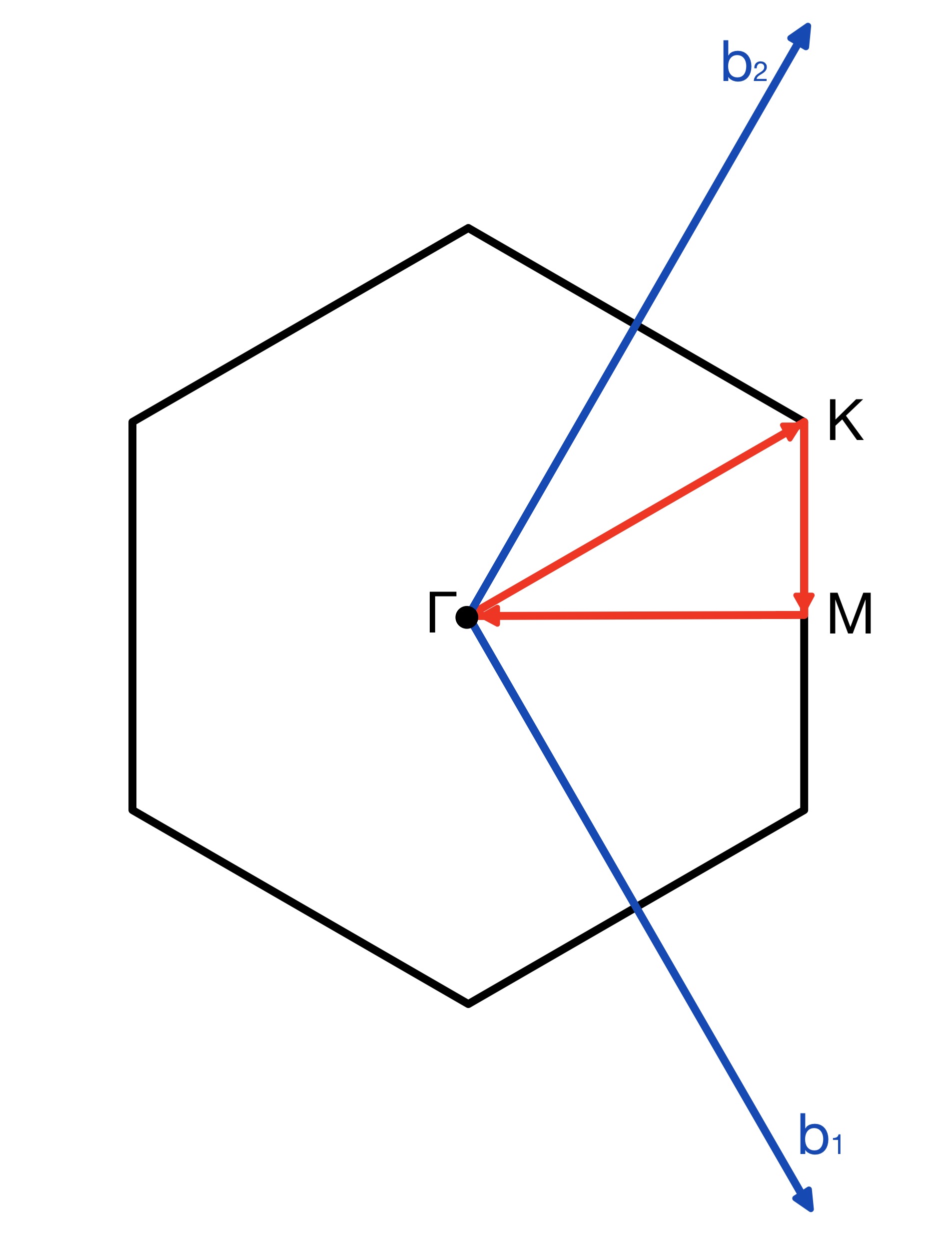}
  \caption{(a)Primitive cell and(b)First Brillouin Zone of hexagonal lattice with high symmetry k-points}
  \label{fig:First Brillouin Zone}
\end{figure}

We calculated and plotted the band structure of WSe$_{2}$ before and after doping using all of the converged values. The high symmetry k-points utilized was $\Gamma$(0, 0, 0), K($1/3$, $2/3$, 0), and M($1/2$, 0, 0) along the hexagonal Brillouin zone as shown in \Cref{fig:First Brillouin Zone}.

We calculated the PDOS for pristine WSe$_{2}$ and transition metal doped WSe$_{2}$. We plotted the 5d orbital of Tungsten, 4p orbital of Selenium, and the d orbital for the dopants to determine whether the localized states above the Fermi level were filled and whether there was strong hybridization. The Fermi level was set to zero.

\subsection{Energy}
In order to calculate the formation energy, we calculated the total energy of WSe$_{2}$ before and after doping. The following formula was used to calculate the formation energy:
\begin{equation}
E_{Formation} = E_{total} - E_{WSe_{2}} - E_{Dopant} + E_{W}
\end{equation}
\\*where E$_{Formation}$, E$_{total}$, E$_{WSe_{2}}$, E$_{Dopant}$, and E$_{W}$ represent the formation energy,  total energy of doped WSe$_{2}$, total energy of pristine WSe$_{2}$, total energy of the dopant, and total energy of a Tungsten atom, respectively.

The adsorption energy was calculated in order to determine the interaction strength between the water molecules and the monolayer by calculating the total energy of WSe$_{2}$ before and after adsorption of H$_{2}$O. The following formula was used to calculate the adsorption energy:
\begin{equation}
E_{Adsorption} = E_{total} - E_{Monolayer} - E_{H_{2}O}
\end{equation}
\\*where E$_{Adsorption}$, E$_{total}$, E$_{Monolayer}$, and E$_{H_{2}O}$ represent the adsorption energy,  total energy of H$_{2}$O adsorbed on the WSe$_{2}$ surface, total energy of pristine or doped WSe$_{2}$ monolayer, and total energy of H$_{2}$O, respectively.

\section{Results}

We first analyzed the electronic calculations of pristine WSe$_{2}$, which were conducted using relaxed and converged values. Next, we demonstrated the effects of Cr, Mo, Ta, and Re doping on WSe$_{2}$. The band structure and PDOS of the doped monolayer to those of the pristine WSe$_{2}$ were compared. Finally, we evaluated the formation energy and water adsorption for each monolayer.

\subsection{Criteria for Water Splitting}
In order for a material to be a photocatalyst, it must satisfy several band gap requirements. The valence band maximum(VBM) must be more negative than the reduction potential of H$^{+}$/H$_{2}$(0eV)\cite{VBM&CBM2019, VBM&CBM2012}. The conduction band minimum(CBM) must be larger than the oxidation potential of O$_{2}$/H$_{2}$O(1.23eV)\cite{VBM&CBM2019, VBM&CBM2012}.
Therefore, the band gap must be larger than 1.23eV. On the other hand, in order for a photocatalyst to efficiently harvest solar energy, the band gap must be smaller than 3.0eV\cite{Splitting2018, GaAs2020}.

\subsection{Photocatalytic Properties of pristine WSe$_{2}$}

\begin{figure}[h]
(a)\includegraphics[width=0.45\linewidth]{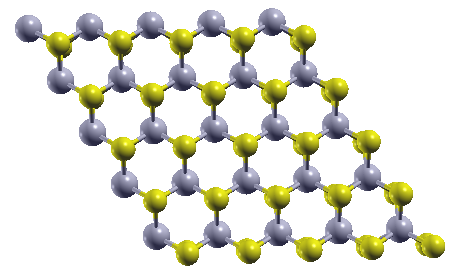}
  \\[\smallskipamount]
(b)\includegraphics[width=0.447\linewidth]{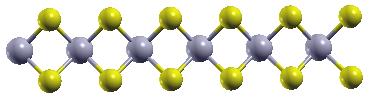}
  \\[\smallskipamount]
(c)\includegraphics[width=0.8\linewidth]{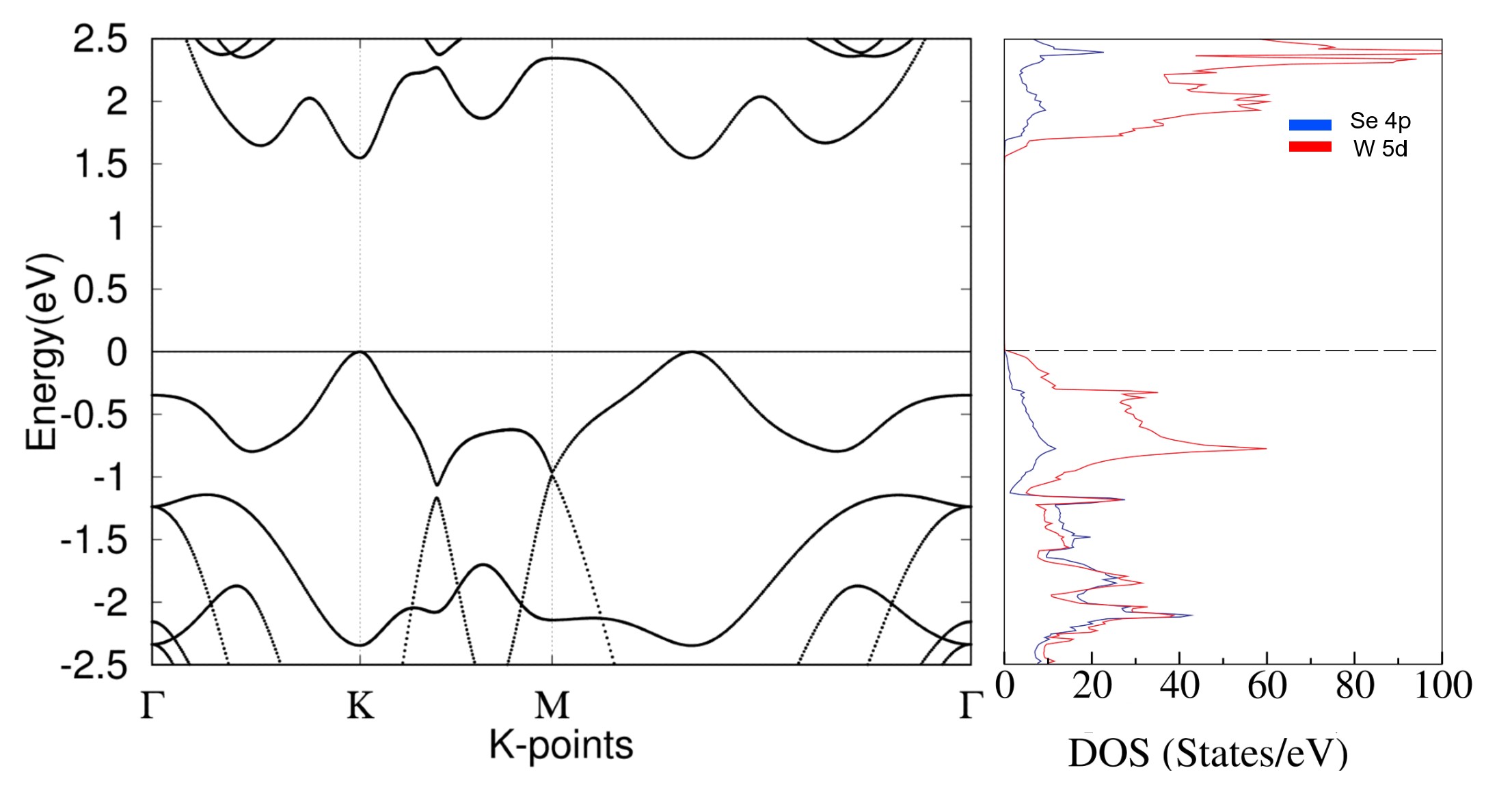}(d)
  
  \caption{(a)Top and(b)side view of the pristine WSe$_{2}$ monolayer. Purple represents the W atoms, and yellow represent the Se atoms.(c)Band Structure and(d)PDOS of pristine WSe$_{2}$. Blue represents Se 4p orbital, and red represents W 5d orbital. The Fermi level is set to 0, which is symbolized as the dotted line.}
  \label{fig:WSe2 crystal structure}
\end{figure}

In this paper, we used the hexagonal monolayer structure of WSe$_{2}$, which is in the P63/mmc space group. As shown in \Cref{fig:WSe2 crystal structure}, the structure consists of sandwich layers Se-W-Se, where the Tungsten layer is enclosed with Selenium. The lattice constant for pristine WSe$_{2}$ is 6.279 Bohr. The bond length between the Selenium and Tungsten is 4.819 Bohr, and the Se-W-Se bond angle is 82.4$^{\circ}$, as listed in \Cref{tab:Table1}. The converged kinetic energy cut off value of WSe$_{2}$ is 23 Hartree. The k-points of 4×4×1 and converged kpoint grid 6×6×1 automatically generated by the Monkhorst-Pack scheme were used for structural optimization and self-consistent calculations, respectively. These values were used to conduct the electronic calculations of WSe$_{2}$. 

As shown in \Cref{fig:WSe2 crystal structure}(c), WSe$_{2}$ has a direct band gap which is slightly higher than 1.5eV. This meets the requirement of a band gap between 1.23eV and 3.0eV. In addition, the band edge positions straddle both the oxidation potential of O$_{2}$/H$_{2}$O and the reduction potential of H$^{+}$/H$_{2}$. These properties establish WSe$_{2}$ as a photocatalyst. The PDOS shown in \Cref{fig:WSe2 crystal structure}(d) indicates that there were no unfilled localized states above the Fermi level. It also shows that the valence band was composed of both the Se 4p and W 5d orbitals, whereas the conduction band was composed mainly of the Se 4p orbital. 

\subsection{Doping Effect on Band Structure and PDOS}
For the doped monolayers, a 3×3×1 supercell was relaxed and utilized for calculations. We doped the monolayer using substitutional doping, where the center tungsten atom was replaced with the dopant. A 2×2×1 kpoint grid and the same kinetic energy cut off were used during the calculations.

\begin{figure}[h]
(a)\includegraphics[width=0.45\linewidth]{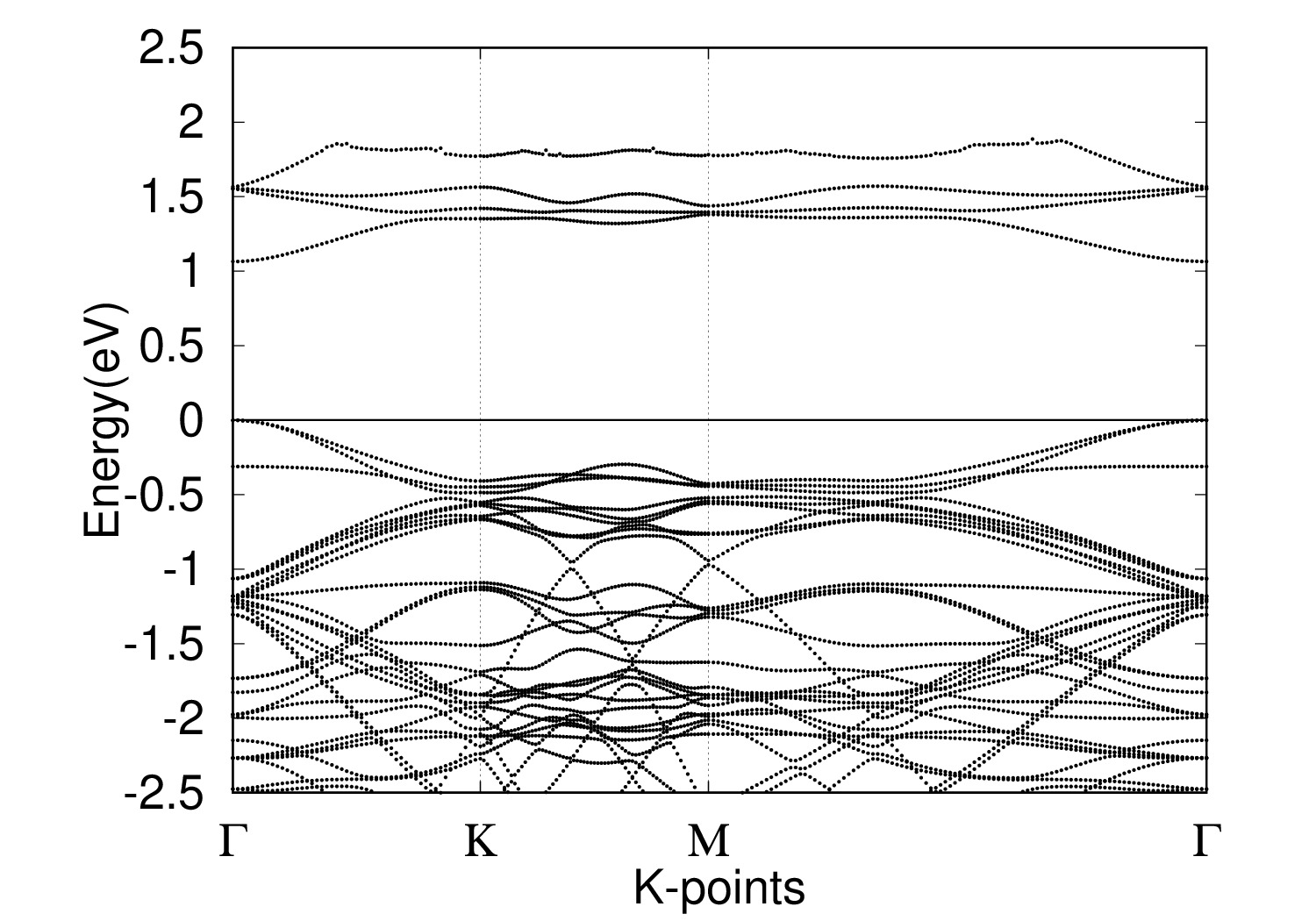}
(b)\includegraphics[width=0.45\linewidth]{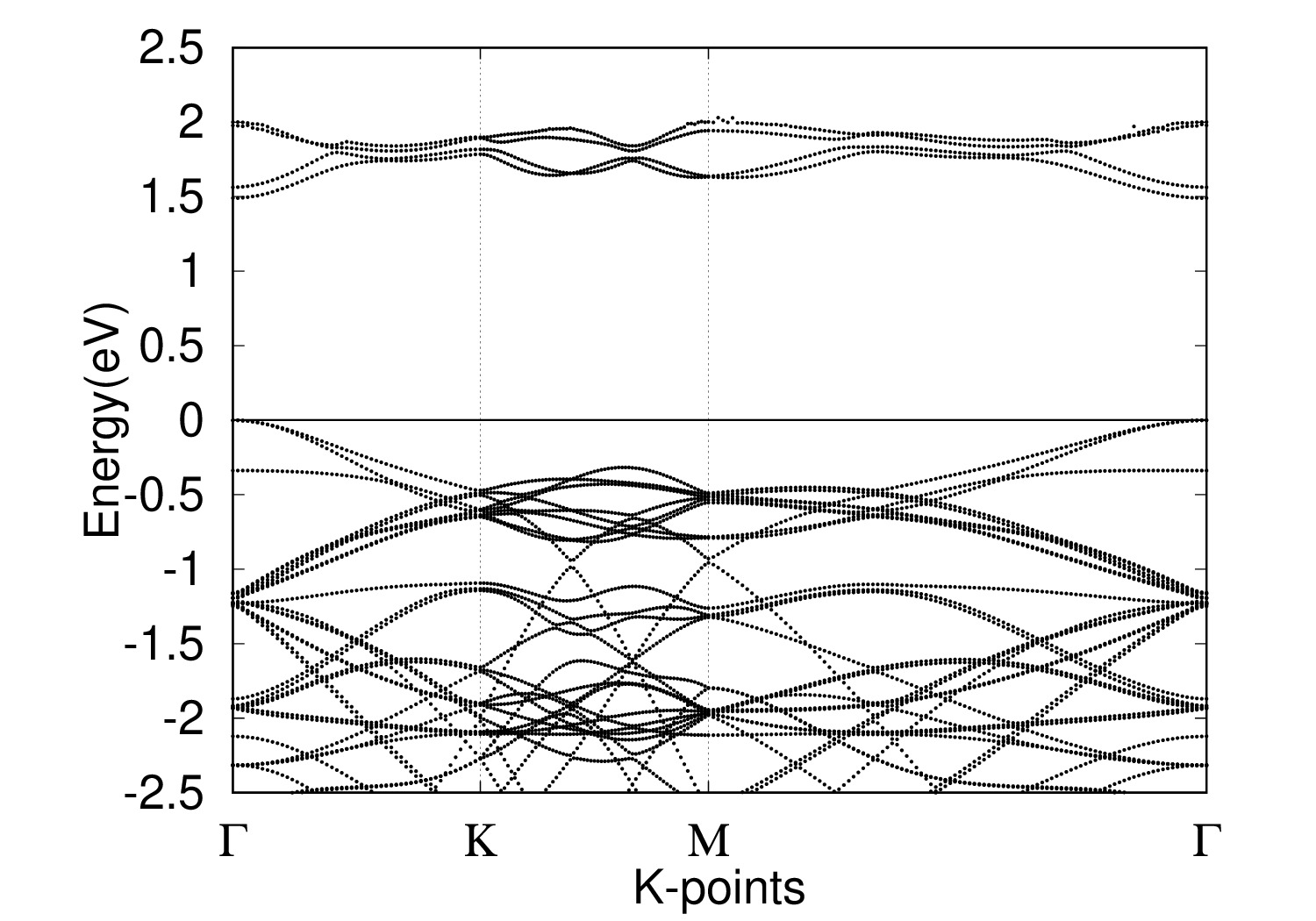}
  \\[\smallskipamount]
(c)\includegraphics[width=0.45\linewidth]{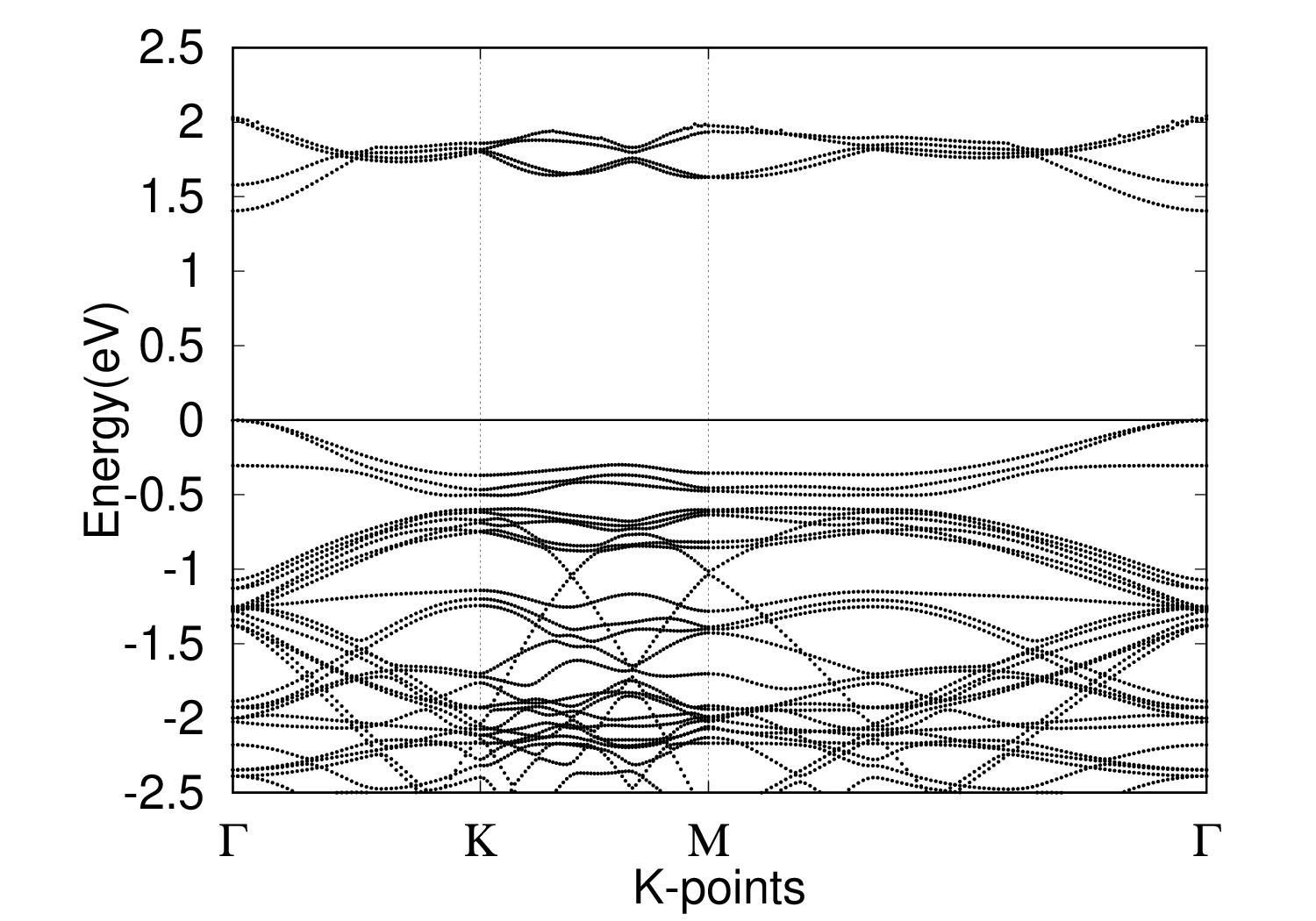}
(d)\includegraphics[width=0.45\linewidth]{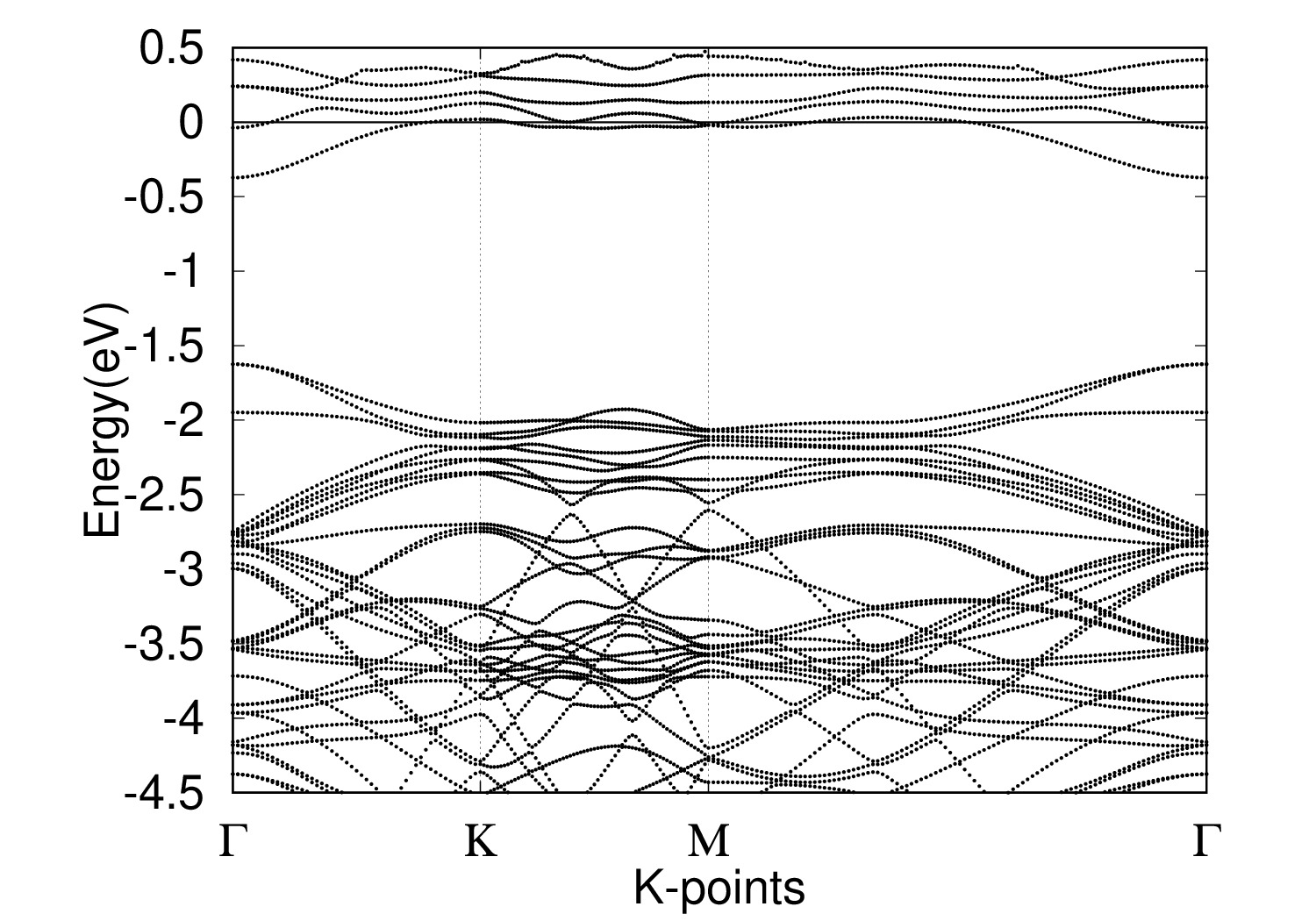}
  \caption{Band structure for(a)Cr,(b)Mo,(c)Ta, and(d)Re doped WSe$_{2}$}
  \label{fig:bandstructure}
\end{figure}

\begin{figure}[h]
(a)\includegraphics[width=0.45\linewidth]{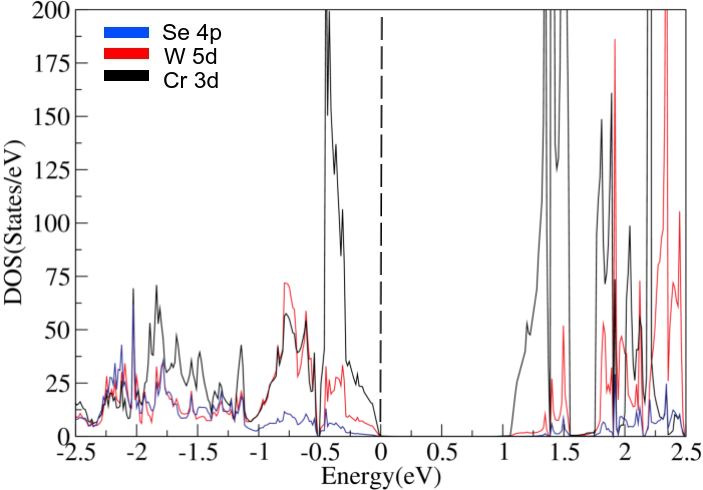}
(b)\includegraphics[width=0.45\linewidth]{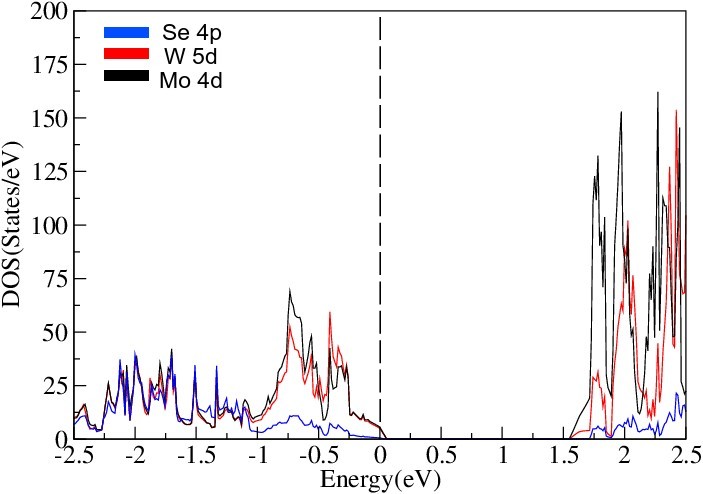}
  \\[\smallskipamount]
(c)\includegraphics[width=0.45\linewidth]{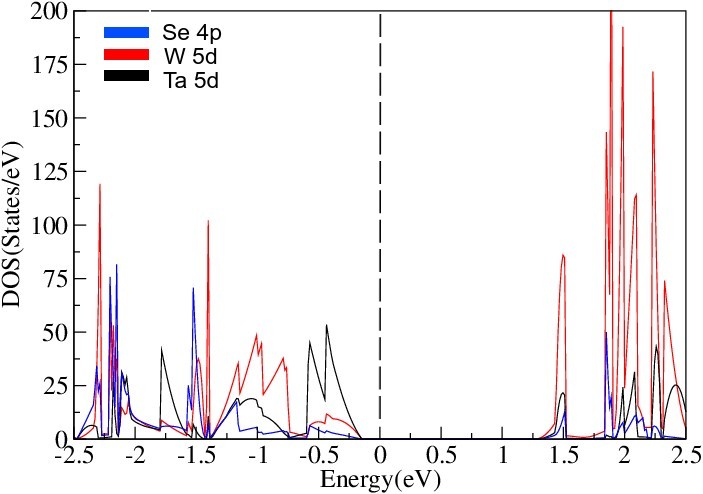}
(d)\includegraphics[width=0.45\linewidth]{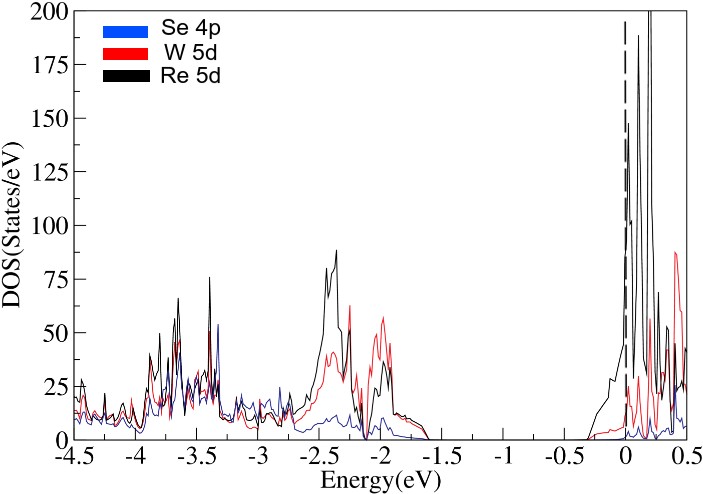}
  \caption{PDOS for(a)Cr,(b)Mo,(c)Ta, and(d)Re doped WSe$_{2}$. Blue, red, and black represent Se 4p, W 5d, and the dopant's d orbital, respectively. The dotted line symbolizes the Fermi level, which is set to 0.}
  \label{fig:PDOS}
\end{figure}

As shown in \Cref{fig:bandstructure}(a) and \Cref{fig:PDOS}(a), the Cr doping negatively impacted the band edge alignments. The Cr dopant caused the shifting of the CBM, as it was the main contributor of the conduction band between 1.1eV and 1.4eV. The new band gap was still direct, however, it was around 1.1eV. The VBM was still at 0, and the CBM was at around 1.1eV, as previously stated, due to the Cr dopant. This is below the oxidation potential of O$_{2}$/H$_{2}$O(1.23eV), therefore, Cr doped WSe$_{2}$ does not meet the criteria to be a photocatalyst. 

In \Cref{fig:bandstructure}(b) and \Cref{fig:PDOS}(b), it is observed that the Mo doping did not change the band gap alignments. The VBM and CBM shifted from the K point to the $\Gamma$ point, however, the band gap was still direct and around 1.5eV, with the VBM at 0eV and the CBM at around 1.5eV, which was very similar to pristine WSe$_{2}$. Also shown in \Cref{fig:PDOS}(b) is the hybridization between the Mo 4d and W 5d orbitals through their simultaneous peaks and troughs, especially between -1 and 0eV. This fits the criteria for Mo doped WSe$_{2}$ to be a photocatalyst, however, it does not increase the number of photons adsorbed because the band gap has not decreased.  

The Ta doping improved the band edge alignments, as shown in \Cref{fig:bandstructure}(c) and \Cref{fig:PDOS}(c). Similar to the Mo doped WSe$_{2}$, both the VBM and CBM shifted to the $\Gamma$ point, thus maintaining the direct band gap. The CBM was lowered by about 0.2eV, and the VBM lowered by about 0.1eV, thus making the new band gap around 1.4eV. Compared to the others, Ta doped WSe$_{2}$ had the smallest band gap without going below 1.23eV. Furthermore, the CBM and VBM straddled the oxidation potential of O$_{2}$/H$_{2}$ and the reduction potential of H$^{+}$/H$_{2}$. It was also observed in \Cref{fig:PDOS}(c) that there were no unfilled localized states above the Fermi level and some hybridization between the Ta 5d orbital and the W 5d orbital. However, the hybridization was not as strong as that of the Mo 4d orbital and the W 5d orbital shown in \Cref{fig:PDOS}(b). Therefore, with the ideal band edge positions and hybridization, Ta doped WSe$_{2}$ is a promising photocatalyst.

As shown in \Cref{fig:bandstructure}(d) and \Cref{fig:PDOS}(d) the Re doping hindered the band edge alignments. The VBM shifted to about -1.6eV, and the CBM shifted to about -0.4eV. Therefore, the band gap is smaller than 1.23eV, and the band edge positions do not qualify Re doped WSe$_{2}$ to be a photocatalyst.

\subsection{Formation Energy}
The formation energy determines how facile it is to perform substitutional doping and how stable the doped monolayer is. If the formation energy is negative, it indicates that the formation is an exothermic process. The more negative the formation energy is, the more facile the doping process is and the more stable the monolayer is. 

As shown in \Cref{tab:Table2}, the formation energies for Mo(-0.448eV) and Ta(-0.922eV) doped WSe$_{2}$ were negative, indicating that they underwent an exothermic process, whereas Cr(0.387eV) and Re(0.845eV) doped WSe$_{2}$ underwent an endothermic process. The formation energies were relatively small, however, out of the 4 different doped monolayers, the Ta doped WSe$_{2}$ monolayer was the most energetically favorable.

\subsection{Water Adsorption}
The water molecule was placed in the hollow site above the dopant because Peng et al. determined that it was the most stable adsorption configuration\cite{CorePaper1}. A negative adsorption energy means that the water molecule can be adsorbed energetically favorably. The more negative the adsorption energy is, the stronger the interaction is between the water molecules and the monolayer.

The water adsorption for each monolayer is listed in the \cref{tab:Table2}. For pristine WSe$_{2}$, the adsorption energy was -0.1184eV, which is relatively weak. This was not improved by the dopants. For the Mo doped WSe$_{2}$ monolayer, the adsorption energy was -0.1174eV, and for the Ta doped WSe$_{2}$ monolayer, surprisingly, the adsorption energy was -0.0439eV. The pristine WSe$_{2}$ and Mo doped WSe$_{2}$ adsorption values were not as negative as we expected, however, they were more negative than the Ta doped WSe$_{2}$ monolayer. This suggests stronger interactions between the water molecules and the pristine and Mo doped WSe$_{2}$ monolayers. Despite the weaker interactions, the adsorption energy values were negative for all three monolayers, meaning that the water molecules can be adsorbed energetically favorably.

\begin{table}[h]

\caption{Bond angles(Se-W-Se), bond lengths(Se-W), Formation energy of transition metal(E$_{f}$), and Adsorption energy of water(E$_{a}$) for pristine and Cr, Mo, Ta, Re doped WSe$_{2}$}
\begin{tabular}{c c c c c}
Configuration&Bond Angle(\degree)&Bond Length(Bohr)&E$_{f}$(eV)&E$_{a}$(eV)\\

\hline
\hline
WSe$_{2}$&82.4&4.819&-&-0.1184\\
WSe$_{2}$(Cr)&82.8&4.819&0.387&-\\
WSe$_{2}$(Mo)&82.3&4.819&-0.448&-0.1174\\
WSe$_{2}$(Ta)&81.9&4.819&-0.922&-0.0439\\
WSe$_{2}$(Re)&82.5&4.819&0.845&-\\

\end{tabular}
\label{tab:Table2}
\end{table}

\section{Discussion}
Photocatalytic water splitting is a promising renewable energy source as an alternative for limited fossil fuels. The effectiveness of the conversion from solar energy to hydrogen fuel relies primarily on the material. Previously, researchers studied different TMDs such as WS$_{2}$\cite{WS2}, MoS$_{2}$\cite{Pan2014, Lolla2020, AnionMoS22017, MoS2bandgap, MoS2VBMCBM}, and PdSe$_{2}$\cite{PdSe22017}. 
These materials perform well in many properties such as strong adsorption stability(WS2) and promising abilities for HER, however, their band gaps are still not ideal. In this paper, we aim to improve upon this band gap by studying a new TMD material, WSe$_{2}$.

Currently, WSe$_{2}$ is used in heterostructure photocatalysts. To our knowledge, this is the first time that pristine monolayer WSe$_{2}$ and doped WSe$_{2}$ have been studied as potential photocatalysts. In particular, our calculations show excellent band gaps and band edge positions, the two most important properties of a photocatalyst. For band gaps, our calculations are 1.5eV for pristine WSe$_{2}$, 1.5eV for Mo doped WSe$_{2}$, and 1.4eV for Ta doped WSe$_{2}$, which are closer to 1.23eV(the ideal band gap which most efficiently harvest solar energy\cite{Splitting2018, GaAs2020}) than previous studies. For example, Long et al(2018) reported a band gap of 2.28eV for PdSe2\cite{PdSe22017}, Ryou et al(2016) reported 1.8eV for MoS2\cite{MoS2bandgap}, and Yao et al(2019) reported 2.36eV for SbP3 and 1.45eV for GaP3\cite{SbP32019}. 

Furthermore, our calculations demonstrated that the band edge positions of pristine, Mo doped, and Ta doped WSe$_{2}$ are an improvement over those of previously reported materials. It is well known that for an ideal photocatalyst, the VBM is the reduction potential of H$^{+}$/H$_{2}$, i.e., 0eV; and the CBM is the oxidation potential of O$_{2}$/H$_{2}$O, i.e., 1.23eV. Previously, Zhang et al reported approximately -0.3eV as the VBM of MoS$_{2}$ and around 2eV as the CBM\cite{MoS2VBMCBM}. In Bui et al's 2015 paper, the VBM for WS$_{2}$ was around -0.2eV and the CBM was about 1.7eV\cite{WS2}. In our paper, for two materials, pristine and Mo doped WSe$_{2}$, the VBM is 0eV and the CBM is 1.52eV. For the third material, Ta doped WSe$_{2}$, the VBM is -0.1eV and the CBM is 1.32eV. For all three materials, the VBM and CBM are much closer to the ideal band edge positions of 0eV and 1.23eV than the previous papers, suggesting a more efficient collection of solar energy.

In addition, as previously discussed, the formation energies for two studied materials, Mo and Ta doped WSe$_{2}$, are negative, indicating a desired exothermic process with stable doped monolayers. Between the Mo doped WSe$_{2}$ and the Ta doped WSe$_{2}$, the Ta doped WSe$_{2}$ monolayer is the most stable. On the other hand, the formation energies for the Cr and Re doped WSe$_{2}$ are positive, suggesting an unstable monolayer. Hence, Cr and Re doped WSe$_{2}$ were dropped in the calculations. 

Despite all the desired properties, one limitation of pristine and doped WSe$_{2}$ is that their performance is less satisfactory for water adsorption. The adsorption energies for pristine WSe$_{2}$, Mo doped WSe$_{2}$, and Ta doped WSe$_{2}$ are -0.1184, -0.1174, and -0.0439 respectively, indicating stable but weak water adsorption. 

Some of the future works can consider calculating spin-orbital coupling, predicting the reaction barrier, and using alternative methods such as the Heyd–Scuseria–Ernzerhof(HSE) exchange–correlation functional and GW approximation(GWA) to calculate the electronic structures, which can then be compared to the PBE exchange–correlation functional method.

\section{Conclusion}

Aiming to improve the band gap of previously reported photocatalytic materials, we studied a new TMD material, WSe$_{2}$, as a monolayer. To our knowledge, this is the first report of using transition metal doped WSe$_{2}$ as potential photocatalysts for photocatalytic water splitting. We used first principles calculations to evaluate the photocatalytic abilities of pristine WSe$_{2}$ as well as Cr, Mo, Ta, and Re doped WSe$_{2}$. Our calculations showed that the band gaps and band edge positions of three of our studied materials(pristine, Mo doped, and Ta doped WSe$_{2}$) are ideal for water splitting. Compared to previous studies of similar materials, our materials demonstrate more desirable band gaps: 1.5eV for pristine WSe$_{2}$ and Mo doped WSe$_{2}$, and 1.4eV for Ta doped WSe$_{2}$, which are closer to the ideal 1.23eV; The band edge positions of our materials are also closer to the reduction potential of H$^{+}$/H$_{2}$ and the oxidation potential of O$_{2}$/H$_{2}$O. In addition, Mo and Ta doped WSe$_{2}$ monolayers undergo an exothermic process, indicating stable monolayers. Of the three selected materials, pristine WSe$_{2}$ exhibits the strongest water adsorption abilities. Our results substantiates pristine, Mo doped, and Cr doped WSe$_{2}$ as potential photocatalysts for water splitting. 
\newpage

\section{Acknowledgements}
The authors would like to acknowledge Dr. Gefei Qian for his technical support.

\section{Author Declarations}
The authors have no conflicts to disclose.

\newpage
\bibliography{references}
\bibliographystyle{unsrt}
\end{document}